\documentclass[a4paper,11pt]{article}
\pdfoutput=1 

\usepackage{jheppub,tensor} 
\usepackage{tikz}
\usepackage[T1]{fontenc} 
\newcommand{\cd}{\bar{\nabla}}
\newcommand{\hhat}{\hat{h}}

\title{\boldmath Effective Non-Minimal Quadratic Gravity and the Classical Cut-Off}

\author[a,b]{Callum L.\ Hunter}
\emailAdd{callum.hunter16@imperial.ac.uk}

\affiliation[a]{Department of Mathematical Sciences\\
Durham University, Lower Mountjoy, Stockton Road, Durham DH1 3LE, UK}
\affiliation[b]{The Blackett Laboratory\\
Imperial College London,
Prince Consort Road, London SW7 2AZ, UK}

\abstract{In this paper we consider Quadratic Gravity as a low energy effective field theory of some unknown UV-complete theory of gravity. Using the spin-2 ghost pathology that occurs in Quadratic Gravity, we derive the maximum value for the classical cut-off of the low energy effective field theory in various configurations. We then add higher order, bottom-up, effective terms. At each order we allow the introduction of non-minimally coupled scalar field terms, and we calculate the vacuum expectation value of the theory under an assumed $\mathbb{Z}_2$ symmetry breaking of the scalar field. This symmetry breaking leads us to introduce a constant in the action which is required to ensure that the physical spin-2 degree of freedom is massless in the flat space approximation, an implicit assumption we make throughout the work. We then consider the case in which we do not fine tune the introduced constant, which leads to an (Anti-)de Sitter solution about which we can then expand. The sign of the chosen cosmological constant is the major factor in determining whether the geometry is de Sitter or Anti-de Sitter. Then expanding about this geometry allows us to increase the theoretical classical cut-off as compared to the flat space case by an amount that is proportional to the free cosmological constant parameter. }

\begin{document} 
\maketitle
\flushbottom

\section{Introduction}
\label{sec:intro}
The unification of general relativity (GR) and quantum field theory (QFT) into a theory of everything is one of the most pressing issues facing modern theoretical physics. A variety of theories have been developed in order to tackle this problem, including String Theories \cite{polchinski1998string1, polchinski1998string2} and Loop Quantum gravity \cite{rovelli2014covariant}. The main issue with uniting the two theories lies in the issue of GR's non-renormalizability at loop level, a fact which is nicely shown in \cite{GRnon-renormexample}. In a typical field theory, one might add higher order counter terms to prevent loop divergences occurring and this can be done for GR. In \cite{stellerenorm} Stelle showed that it is possible to make GR renormalizable to all orders of perturbation theory by adding the quadratic terms $\alpha R^2+\beta R_{\mu\nu}R^{\mu\nu}$. Unfortunately, we have to pay a price for this renormalizability which is the introduction of new massive modes including a spin-0 massive mode and a spin-2 ghost massive mode \cite{StelleHighDeriv,massivemasslessandghost, Ghostinmodgrav,2005PhLB..608..189N,2005JCAP...03..008C}. The spin-0 mode is not too much of an issue, as outlined in \cite{massivemasslessandghost} the spin-0 mode's coupling may be so small that its influence is not detectable and hence its existence cannot be ruled out from experiment. Alternatively, the mass of the spin-0 mode may be so large that it contributes very little to the low energy behaviour that is typically investigated by experiment. In fact, new fields such as the scalar field, or spin-0 field, are not new in the search for a theory of quantum gravity \cite{Shojai:1999jh,Shojai:2000us}. The central point here is that the spin-0 is not \textit{pathological}, it can exist within the theory and the only issues it may cause are some departures from experimental results, such as fifth force effects, however this can be dealt with by using sufficiently small couplings, or a sufficiently large mass. Since the spin-2 mode is a ghost, it \textit{is} a pathology of the theory and causes unbounded negative energy in the quantum theory \cite{GhostNotes}, and instabilities in the classical theory.

This fact, however, has not prevented further investigation into so-called modified theories of gravity. A rich and varied field of work has been completed on theories of varying structure, including $f(R)$ gravity \cite{fRreview,thomasf(r)} and more general functions of the curvature tensors \cite{generalmodified, Lovelock}. All of these theories aim to solve issues in cosmology or high energy physics, but many contain the pathology of the spin-2 ghost mode.

Another type of modified gravity introduces a scalar field which non-minimally couples to gravity, affecting the dynamics of the theory. The most famous example of this theory type is the Brans-Dicke theory \cite{bransdicke} which introduces a non-minimally coupled scalar field which ultimately leads to Newton's constant $G$ becoming a function of the scalar field \cite{fujii2003scalar}. There are further examples of such theories including Einstein-dilaton-Gauss-Bonnet gravity \cite{EdGB, EdGBBH} and its scalar generalisation \cite{EGBS}. These theories have mostly been investigated in the context of black hole physics, and the violation of the no-hair theorems \cite{no-hair1,no-hair2, 2020arXiv201010312H}. However, owing to the Gauss-Bonnet's quadratic structure, these too will contain the ghost-modes previously highlighted.

\begin{figure}
    \centering
    \includegraphics[width=0.9\linewidth]{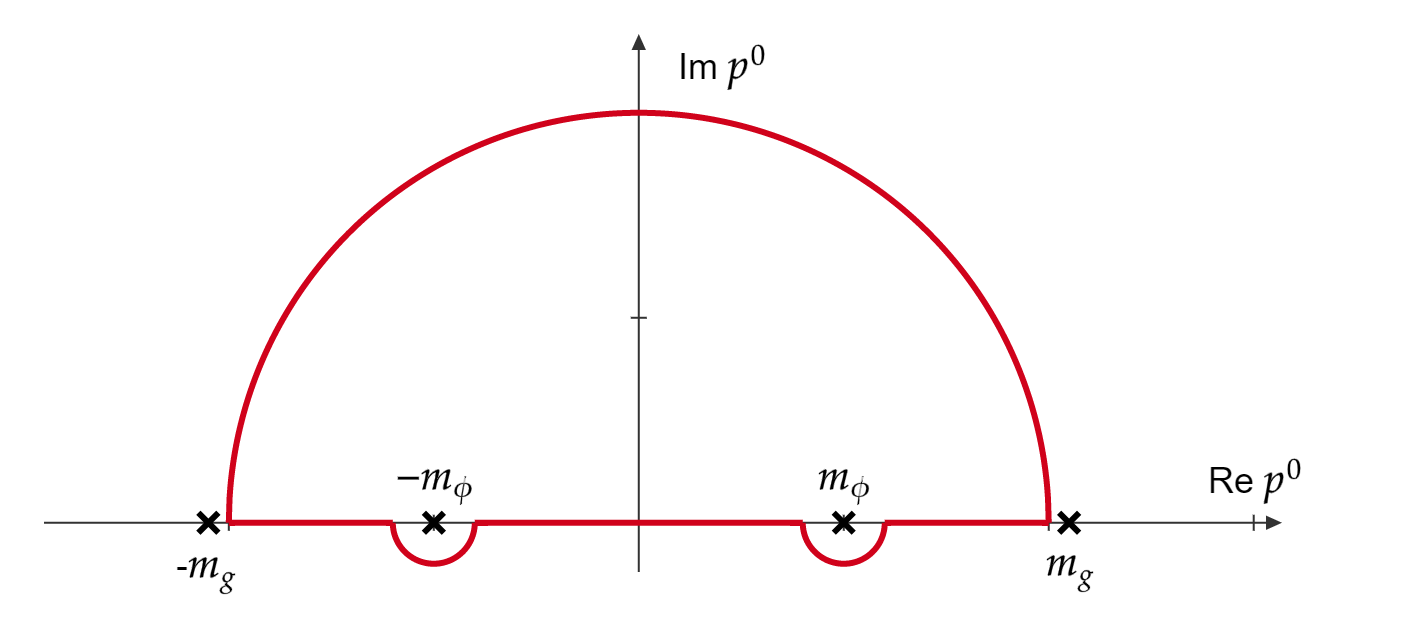}
    \caption{The contour for the solution to a classical field theory with a particle of mass squared $m_\phi^2$ and a ghost particle of absolute mass squared $m_g^2$. The position of the $m_g$ may be on the real or imaginary axis depending on whether the ghost mode is a tachyon or not.}
    \label{fig1}
\end{figure}

It now seems as though all hope is lost. However, if we treat the theory as an effective field theory (EFT), then we can set the cut-off of the effective field theory below the mass of the ghost mode and hence prevent it from being a dynamical field\footnote{We should point out that this assumption only really holds in the bottom-up EFT case, as in the top down case one would have to integrate out the massive mode in order to find the true dynamics of the theory.}, see Figure \ref{fig1}. The treatment of GR and GR-like theories as EFTs is not new \cite{Donoghue1,Donoghue2,Solomon:2017nlh} and can be an immensely powerful tool in dealing with pathologies in the kind of theories we wish to consider. There are also examples of gravity coupled to other effective theories such as electrodynamics \cite{2016CQGra..33i5008I}. In the case of black holes, this kind of theory leads to interesting conclusions including the fact that extremal black holes become self-repulsive in certain limits \cite{2019PhRvL.123y1103B}. 

Our work will take a \textit{bottom-up} approach, as \cite{2019PhRvL.123y1103B} does with electrodynamics. In this scenario one begins with the low energy effective field theory (LEEFT), such as GR or Quadratic Gravity, and then proceeds to add all possible terms that obey the necessary symmetry requirements of the theory. This is done order-by-order in the mass dimensions of the cut-off scale, $\Lambda$. For example, the LEEFT of GR is given by the Einstein-Hilbert action,
\begin{equation}\label{1.1}
    S_{EH}=\int d^4x \sqrt{-g}M_P^2(R),
\end{equation}
where $M_P^2$ is the Planck mass and $R$ has dimensions of mass squared. Now at the next order, one may add $\alpha R^2+\beta R_{\mu\nu}R^{\mu\nu}$, but this term does not require the $M_P^2$ term to be of the correct dimension,
\begin{equation}\label{1.2}
    S_2=\int d^4x \sqrt{-g}(M_P^2R+\alpha R^2+\beta R_{\mu\nu}R^{\mu\nu}).
\end{equation}
Now, the contribution from the quadratic part of the action is $\mathcal{O}(M_P^2)$ smaller than the usual Einstein-Hilbert contribution, which is very small indeed. One can continue in this manner and add on the next terms in the sequence which will yield,
\begin{equation}\label{1.3}
    S_3=\int d^4x \sqrt{-g}(M_P^2R+\alpha R^2+\beta R_{\mu\nu}R^{\mu\nu}+\tfrac{\alpha_1}{M_P^2}R\Box R+...),
\end{equation}
where $M_P$ would be the cut-off mass in this example. We assume that $\Lambda=M_P$ is a large quantity and hence only a finite number of terms are required to actually capture a large amount of the physics. It may seem initially troublesome that we have a ghost mode in the theory, as it may appear that no matter what energy we work at we shall also have a vacuum that is unstable due to the negative energy of the ghost. However, we make an assumption about the nature of the EFT which resolves this issue.

\paragraph{Assumption} Here we state the assumption about the classical ghost, based on logic outlined in \cite{massivemasslessandghost,fontanini2011higher}. We assume that above the cut-off mass new interactions enter into the theory which save us from the ghost mode. The cut-off mass can then at most be assumed to be around the ghost mode mass itself \cite{fontanini2011higher, 2015EJPh...36a5009S}, as it is this mode which is troublesome. Hence we can define the contour of the solution to our theory in the normal manner, but with non-infinite limits on the integral as in Figure \ref{fig1}.

\medskip

\noindent As our basis, we shall use Quadratic gravity as our LEEFT, and so our first order action will be the action given in \eqref{1.2}. It is worth pausing here and briefly considering the degrees of freedom in the action \eqref{1.2}. We do not calculate the degrees of freedom here, rather we will quote the results for the degrees of freedom of Quadratic Gravity. This is because only the degrees of freedom appearing in the LEEFT contribute to the spectrum. There are three modes present in the action \eqref{1.2}, these are: the spin-0 gravitational scalar, the spin-2 massless mode corresponding to the graviton and the spin-2 ghost particle \cite{2018FrP.....6...77S}. The first of these carries a single degree of freedom, the second carries two degrees of freedom corresponding to the two polarizations of the gravitational waves and finally the ghost mode carries six degrees of freedom, this draw parallels with the Boulware-Desser Ghost \cite{2014LRR....17....7D}. This means the spin-2 degrees of freedom (real and ghost) add together to give the expected eight \cite{2019PhRvD..99f4039H}. Thus, in total there are nine degrees of freedom in the theory.

In this work we shall assume that Quadratic gravity is only valid up to some scale $\Lambda$, and that the EFT terms allow us to change this cut-off. The reason for this choice of action is somewhat arbitrary, however in picking \eqref{1.2} as our LEEFT we have the ghost mode as a fixed mode in the spectrum. If one had chosen \eqref{1.1} as the LEEFT, then despite the higher order terms, no ghost modes would be present in the EFT. This fact follows from principle the that since the field theory is effective, the new modes that would appear in the spectrum due to higher order terms cannot appear since these new propagator terms must be treat \textit{pertubatively} \cite{spintwocoupling}. For example, taking GR as the LEEFT, one may obtain an effective propagator of the form,
\begin{equation}\label{1.4}
    G(k)\propto \frac{1}{k^2+\Gamma k^4}\propto\frac{1}{k^2}-\frac{1}{k^2+\Gamma},
\end{equation}
which introduces a new mode. However, this does not treat the effective terms as perturbations, but rather treats them on the same footing as the LEEFT \cite{BurgessQuantumGrav}. In order to remain consistent with the nature of the EFT we must treat the higher order terms as perturbations from the LEEFT, and hence \eqref{1.4} becomes \cite{BurgessQuantumGrav},
\begin{equation}
    G(k)\propto\frac{1}{k^2}\bigg(1-\frac{ k^2}{\Gamma}+...\bigg)
\end{equation}
As such, by taking GR as the LEEFT, one always ends up with a massless spin-2 mode as the only propagating mode, albeit with modified dynamics. This is not an obvious reason to pick \eqref{1.2} over \eqref{1.1}, however \eqref{1.1} does not give us any pathology to indicate that the classical cut-off departs from the Planck Mass. In \eqref{1.2} we have the ghost mode, which serves as a classical upper-limit on the cut-off of the EFT. Hence we can use this ghost mode to derive an upper-most cut-off mass without resorting to quantum calculations as in the GR case \cite{GRcutoff}. This shall be the general methodology adopted in this work.

We shall also include a non-minimally coupled scalar field in the action in analogy with the Higgs doublets in \cite{nonmin1,nonmin2}. We do this as a toy model for a more complex analysis, such as the one carried out in \cite{2003AnPhy.305...96A}, concerning the derivation of massive gravitons via a symmetry breaking procedure \cite{2007arXiv0708.3184T}. These non-minimal terms will not affect the situation in which the scalar field has a zero vacuum expectation value, however in the case of $\mathbb{Z}_2$ symmetry breaking, it shall contribute directly to the cut-off mass of the theory. The symmetry breaking introduces some issues such as an induced cosmological constant and a coupling of fields at quadratic order. Dealing with the coupled fields is rather complex as they are a general consequence of having non-minimally coupled terms, such as $\phi^2 R$ \cite{2019JHEP...10..163R}, in the action. The spin-0 fields are not important to the arguments presented in this paper and so we shall not present the decoupling. However the induced cosmological constant will have to be dealt with in Section \ref{S3} because, if we take the expansion about Minkowski space, despite the constant, it will force the spin-2 field to become massive \cite{scharf2016gauge}. This is a situation we wish to avoid as we shall assume throughout that our spin-2 physical mode is massless. If we keep the constant, but force the spin-2 field to be massless, then we can no longer expand about flat space as it is no longer a solution to the theory and we would have to expand about (Anti-)de Sitter space. In order to avoid this issue, we shall add a cosmological constant on to the original theory in Section \ref{S3}, so that once SSB takes places, there are no cosmological constant-like terms. This cosmological constant will have to be fined tuned in the Minkowski case.

This paper is organised in the following manner. In Section \ref{S2} we shall detail the effective action we intend to consider and calculate the perturbative vacuum expectation value of the scalar field under symmetry breaking in Minkowski space. In Section \ref{S3} we shall explore the flat background case, as this is the regime in which mode masses can be defined without effects due to curvature. We shall begin by exploring the case of no vacuum expectation value for the scalar field, that is with the $\mathbb{Z}_2$ symmetry in tact, and derive the masses of the mode spectrum. Then we shall allow $\mathbb{Z}_2$ breaking to occur. In Section \ref{S4} we shall consider the case whereby we do not include the fixed preemptive cosmological constant under SSB and hence calculate the curvature of (Anti-)de Sitter space required to keep the spin-2 field massless. Finally in Section \ref{S5} we shall draw our conclusions and detail some further thoughts on the subject.

\paragraph{NB:} In this work we use the conventions $g_{\mu\nu}=(-,+,+,+)$, $\tensor{R}{^\mu_{\nu\rho\sigma}}=\tensor{\Gamma}{^\mu_{\nu\sigma,\rho}}-\tensor{\Gamma}{^\mu_{\nu\rho,\sigma}}+...$, $R_{\mu\nu}=\tensor{R}{^\rho_{\mu\rho\nu}}$ and that all indices run from 0 to 3. When referring to the field $\phi$ we use `scalar field', however when referring to the gravitational scalar degrees of freedom or the scalar fields in general, we use `spin-0 field(s)'. Throughout, we use `mode' to refer to the fields. We do this because we are not treating the theory as a quantum field theory and so the notion of `particle' is not quite defined. We have stuck to use of the word `mode' to ensure that it is clear we are dealing with a \textit{classical} field theory. Finally, we use Quadratic gravity, with a capital Q, to refer to the theory investigated by Stelle and a lower case quadratic gravity to refer to the quadratic expansion in the perturbation of the metric, denoted by $h_{\mu\nu}$.

\section{The Action and SSB Value}\label{S2}
In this section we shall introduce the dimension-6 EFT and derive the perturbed vacuum expectation value of the scalar field in the flat space limit. The assumed symmetry of the scalar field is the discrete $\mathbb{Z}_2$ symmetry which we shall then break by allowing the scalar field to obtain a constant vacuum expectation value. The action we use in this paper is based upon the Quadratic gravity given in \eqref{1.2}, however we work with the most general non-minimally coupled Quadratic gravity action. This involves adding a $\xi_1\phi^2R$ term to \eqref{1.2}. This extra scalar adds another degree of freedom giving ten degrees of freedom in total. The action, without cosmological constant, is then given by,
\begin{equation}
\begin{aligned}\label{2.1}
    S=\int d^4x \sqrt{-g}\bigg[&M_P^2R-\frac{1}{2}\partial_\mu\phi\partial^\mu\phi-\frac{m^2}{2}\phi^2-\frac{\lambda}{4}\phi^4+\alpha_1 R^2+\alpha_2R_{\mu\nu}R^{\mu\nu}\\
    &+\frac{\alpha_3}{\Lambda^2}R\Box R+\frac{\alpha_4}{\Lambda^2}R_{\mu\nu}\Box R^{\mu\nu}+\frac{1}{\Lambda^2}(\beta_1(\partial_\mu\phi\partial^\mu\phi)\phi^2+\beta_2\phi^3\Box\phi\\
    &+\beta_3(\Box\phi)^2+\beta_4\phi^6)+\xi_1\phi^2R+\frac{1}{\Lambda^2}(\xi_2\partial^\mu\phi\partial^\nu\phi R_{\mu\nu}\\
    &+\xi_3\partial_\mu\phi\partial^\mu\phi R+\xi_4 \phi^2R^2+\xi_5\phi^2 R_{\mu\nu}R^{\mu\nu}+\xi_6\phi^4R\\
    &+\xi_7\phi^2R_{\mu\nu\rho\sigma}R^{\mu\nu\rho\sigma})\bigg].
    \end{aligned}
\end{equation}
In our notation we let $\{\alpha_i\}$ be couplings for curvature terms, $\{\beta_i\}$ the couplings for scalar terms and $\{\xi_i\}$ the couplings for non-minimally coupled terms. The curvature operators $R\Box R$ and $R_{\mu\nu}\Box R^{\mu\nu}$ are two of the possible ten unique combinations given at dimension-6 in the curvature only operators \cite{dimsixops}. We do not include the other eight since they are cubed in the curvature terms and hence they contribute at $\mathcal{O}(h^3)$ in flat space. Hence this action gives the most general non-minimal EFT that will contribute terms to quadratic order in $h_{\mu\nu}$ in the flat space expansion\footnote{It should be noted that this is the most general EFT provided we assume a theory without torsion, for details on generalising to a case with torsion see \cite{Li_2018} and references therein.}. In Section \ref{S4} we investigate the case of constantly curved space to $\mathcal{O}(1)$ in the cut-off mass and quadratic order in the perturbations. If we investigated to $\mathcal{O}(\Lambda^{-2})$ in the curved space situation then we would have to include the quadratic $h_{\mu\nu}$ terms that result from the dimension-6 operators of the form $R^3$, $RR_{\mu\nu}R^{\mu\nu}$ etc. However, this adds a great deal of complexity and will add very little to the final conclusion. In order to work with \eqref{1.2} as the LEEFT with only $\mathcal{O}(\Lambda^{-2})$ and higher order non-minimal terms, we can set $\xi_1=0$ in our results. For the EFT of \eqref{1.2} without the non-minimal couplings, we simply set $\xi_i=0.$

The $\mathbb{Z}_2$ symmetry leads to the two cases that we shall investigate in Section \ref{S3}. In the first we shall look at a trivial background solution given by $\phi_0=0$ and $g_{\mu\nu}=\eta_{\mu\nu}$; in the second case we shall give the scalar field a vacuum expectation value by spontaneously breaking the scalar $\mathbb{Z}_2$ symmetry, however we shall have to modify \eqref{2.1} in order to avoid a massive spin-2 physical mode in flat space. This potential mass term results from the induced constant which appears due to the vacuum expectation value of the scalar field. As a result, we add a preemptive cosmological constant to \eqref{2.1} in order to ensure we have a massless spin-2 mode. The method will become clear in Section \ref{S3}.

We shall work to only second order in the perturbations about the background solutions since this is all we require in order to determine the masses of the fields. The first case leads to a natural decoupling at quadratic order between the scalar field and the gravitational dynamics such that we have four independent fields evolving on a flat background. In the second case, we shall find that the scalar perturbations and the spin-0 gravitational perturbations couple at the quadratic order. This occurs due to the nature of the non-minimally coupled terms that entered the action. Furthermore, in the second case we shall induce a cosmological constant given in terms of the couplings and scalar mass, this ensures a massless spin-2 physical mode. 

In order to show that $\phi_0=v$ and $g_{\mu\nu}=\eta_{\mu\nu}$ is a valid solution about which we can perturb the action, we shall require the equations of motion. The scalar equation of motion resulting from \eqref{2.1} is given by,
\begin{equation}\label{2.2}
    \begin{aligned}
    &\Box\phi-m^2\phi-\lambda\phi^3+\frac{2\beta_1}{\Lambda^2}(\phi^2\Box\phi+\phi\partial_\mu\phi\partial^\mu\phi)+\frac{6\beta_2}{\Lambda^2}\phi^2\Box\phi\\
    &+\frac{2\beta_3}{\Lambda^2}\Box^2\phi+\frac{6\beta_4}{\Lambda^2}\phi^5+2\xi_1\phi R+\frac{2\xi_2}{\Lambda^2}\partial_\nu(\partial_\mu\phi R^{\mu\nu})\\
    &+\frac{2\xi_3}{\Lambda^2}\partial_{\mu}(\partial^\mu\phi R)+\frac{2\xi_4}{\Lambda^2}\phi R^2+\frac{2\xi_5}{\Lambda^2}\phi R_{\mu\nu}R^{\mu\nu}+\frac{4\xi_6}{\Lambda^2}\phi^3R\\
    &+2\xi_7\phi R_{\mu\nu\rho\sigma}R^{\mu\nu\rho\sigma}=0,
    \end{aligned}
\end{equation}
from which it is obvious that $\phi_0=0$ is a solution, however for $\phi_0=v\neq0$ and $g_{\mu\nu}=\eta_{\mu\nu}$ we shall require the gravitational field equations. These equations are rather cumbersome and hence we shall not display them here. However, they do admit a $\phi_0=v$ and $g_{\mu\nu}=\eta_{\mu\nu}$ solution provided $v$ is given by a specified constant. An example of this is more explicitly shown in Section \ref{S4} for the constantly curved space case. In order for $\phi$ to acquire a vacuum expectation value we allow $m^2<0$ and send $-m^2\rightarrow+\mu^2$. Now taking \eqref{2.2}, along with the gravitational field equations, we find that $v$ must satisfy,
\begin{equation}\label{2.3}
    \mu^2v-\lambda v^3+\frac{6\beta_4}{\Lambda^2}v^5=0,
\end{equation}
which can be solved perturbatively in orders of $\Lambda^{-2}$. In order to do this we solve the $\mathcal{O}(1)$ equation first which gives,
\begin{equation}\label{2.4}
    v_{(1)}^2=\frac{\mu^2}{\lambda},
\end{equation}
we can then evaluate the $\mathcal{O}(\Lambda^{-2})$ term on this solution and solve again for $v^2$ which gives,
\begin{equation}\label{2.5}
v_{(2)}^2=\frac{\mu^2}{\lambda}\bigg(1+\frac{6\beta_4\mu^2}{\Lambda^2\lambda^2}\bigg),    
\end{equation}
which are the expectation values we shall use in Section \ref{S3}. Hence we now have that $g_{\mu\nu}=\eta_{\mu\nu}$ and $\phi_0=0,v$ are solutions to \eqref{2.2} and the gravitational field equations. We now turn our attention to the pedagogical flat space limit.

\section{The Weak Field Limit and Quadratic Action}\label{S3}

The main aim we are working towards here is to derive the upper classical limit for the cut-off mass, $\Lambda$, via the spin-2 ghost mass. In order to do this we shall substitute,
\begin{align}
    g_{\mu\nu}&=\bar{g}_{\mu\nu}+h_{\mu\nu},\label{3.1}\\
    \phi&=\phi_0+\chi\label{3.2},
\end{align}
into \eqref{2.1}, where $|\chi|,|h_{\mu\nu}|\ll1$ and $\phi_0=v$ is the vacuum expectation value. In this section we work in the flat limit and so the background metric is the Minkowski metric, that is $\bar{g}_{\mu\nu}=\eta_{\mu\nu}$. Then we shall calculate the mass of the spin-2 ghost mode and then set the limit of the cut-off mass as less than the spin-2 ghost mode mass, as per the assumption in Section \ref{sec:intro}. In order to calculate the ghost mode mass we shall iteratively solve for the mass order-by-order in $\Lambda^{-2}$. As we shall see, when we add on higher order terms we can change the maximum value for the cut-off mass. We may increase or decrease this cut-off mass depending on the values we choose for the effective couplings constants. We shall also see that breaking the scalar symmetry allows us to raise the cut-off value such that $\Lambda^2\propto\mu^2$.

\subsection{The Case of $v=0$}
In this section we shall substitute \eqref{3.1} and \eqref{3.2} into \eqref{2.1} with $v=0$. This will then yield an effective action which is quadratic in the fields, and from this we can then extract the masses of the modes by transforming into momentum space. In the following $\hat{h}=\partial_\mu\partial_\nu h^{\mu\nu}$ and $\hat{h}^\nu=\partial_\mu h^{\mu\nu}$, and we shall use this notation throughout where the full expression is long\footnote{In Section \ref{S4} we shall send $\partial\rightarrow\nabla$ in order to take care of the fact we shall be working in curved space.}. Upon substitution, the effective action is then given by,
\begin{equation}\label{3.3}
    \begin{aligned}
    \mathcal{S}=\int d^4x& \bigg(\frac{1}{2}\chi\Box\chi-\frac{m^2}{2}\chi^2+\frac{2\beta_3}{\Lambda^2}\chi\Box^2\chi\bigg)+\frac{M_P^2}{2}\bigg(h_{\mu\nu}\Box h^{\mu\nu}-h\Box h+2h\hat{h}+\hat{h}^\nu\hat{h}_{\nu}\bigg)\\
    &+2\alpha_1\bigg(\hat{h}(\hat{h}-2\Box h)+h\Box^2h\bigg)+\frac{\alpha_2}{2}\bigg(h_{\mu\nu}\Box^2h^{\mu\nu}+h\Box^2h+2\hat{h}(\hat{h}-\Box h)\\
    &+2\hat{h}^\nu\Box\hat{h}_{\nu}\bigg)+\frac{2\alpha_3}{\Lambda^2}\bigg(\Box\hat{h}(\hat{h}-2\Box h)-h\Box^3h\bigg)+\frac{\alpha_4}{2\Lambda^2}\bigg(2\Box\hat{h}(\hat{h}-\Box h)\\
    &+2\hat{h}^\nu\Box^2\hat{h}_\nu+h\Box^3h+h_{\mu\nu}\Box^3h^{\mu\nu}\bigg),
    \end{aligned}
\end{equation}
to quadratic order in the metric perturbations. Now let us turn our initial attention to the scalar field equation. To first order we have have the usual Klein-Gordon equation of motion,
\begin{equation}\label{3.4}
    \Box\chi-m^2\chi=0,
\end{equation}
which gives the scalar field mass as $m_\chi^2=m^2$ as expected. Now to next order the equation of motion is given by,
\begin{equation}\label{3.5}
    \frac{4\beta_3}{\Lambda^2}\Box^2\chi+\Box\chi-m^2\chi=0\rightarrow\Box\chi-\bigg(m^2-\frac{4\beta_3m^2}{\Lambda^2}\bigg)\chi=0,
\end{equation}
where we have assumed that $\Box^2\chi=\Box(\Box\chi)=m^2\Box\chi$. This is a sensible assumption provided we perturbatively solve for the mass of the mode in order of $\frac{1}{\Lambda^2}$. To this order the mass of the scalar mode is then given by,
\begin{equation}\label{3.6}
    m_\chi^2=m^2\bigg(1-\frac{4\beta_3m^2}{\Lambda^2}\bigg).
\end{equation}
We could have transformed into momentum space and then solved for the mass in that manner, however such a treatment is not required for the scalar field. It will make things more explicit for the gravitational side of things which we turn to now. In the following we shall roughly follow the conventions set out in \cite{massivemasslessandghost} and hence we shall rewrite the perturbations as,
\begin{equation}\label{3.7}
    h_{\mu\nu}=\Bar{h}_{\mu\nu}+\eta_{\mu\nu}\theta,
\end{equation}
where $\theta$ is the trace of the perturbation, thus leaving $\bar{h}_{\mu\nu}$ traceless. Then we shall apply the gauge condition $\partial_\mu\Bar{h}^{\mu\nu}=0$, and in doing so $h_{\mu\nu}=\Bar{h}_{\mu\nu}+\eta_{\mu\nu}\theta$ and $h=4\theta$. In applying this gauge fixing, the effective action for the gravitational perturbations then becomes,
\begin{equation}\label{3.8}
    \begin{aligned}
    \mathcal{S}_{grav}=\int d^4x \bigg[&\frac{M_P^2}{2}\Bar{h}_{\mu\nu}\Box\Bar{h}^{\mu\nu}+\frac{\alpha_2}{2}\Bar{h}_{\mu\nu}\Box^2\Bar{h}^{\mu\nu}+\frac{\alpha_4}{2\Lambda^2}\Bar{h}_{\mu\nu}\Box^3\Bar{h}^{\mu\nu}\\
    &-3M_P^2\theta\Box\theta+(18\alpha_1+6\alpha_2)\theta\Box^2\theta+\frac{18\alpha_3+6\alpha_4}{\Lambda^2}\theta\Box^3\theta\bigg],
    \end{aligned}
\end{equation}
the resulting equations of motion are then easy to write down and are given by,
\begin{align}
    M_P^2\Box\Bar{h}_{\mu\nu}+\alpha_2\Box^2\Bar{h}_{\mu\nu}+\frac{\alpha_4}{\Lambda^2}\Box^3\Bar{h}_{\mu\nu}&=0\label{3.9}\\
    M_P^2\Box\theta-(6\alpha_1+2\alpha_2)\Box^2\theta-\frac{6\alpha_3+2\alpha_4}{\Lambda^2}\Box^3\theta&=0\label{3.10}.
\end{align}
Now to make the mode spectrum more clear we shall transform into momentum space by making the replacements $\Box\Bar{h}_{\mu\nu}\rightarrow -k^2\Bar{h}_{\mu\nu}$ and $\Box\theta\rightarrow -k^2\theta$. It may appear, from first glance, that \eqref{3.10} implies the existence of a massless spin-0 mode, however this is not the case and results from extra gauge freedom in analogy with GR. We can easily show that the only mode that exists is a spin-0 massive mode by looking at the equations of motion of the LEEFT, as the LEEFT defines the mode spectrum, see Appendix \ref{App1}. We shall deal with \eqref{3.9} first. The methodology that we shall follow for the higher orders is the standard procedure to perturbatively solve for the mass. The general procedure is to solve the first order equation and then evaluate the next order term on this solution and solve the previous order. That is, if we solve our $\mathcal{O}(1)$ LEEFT equations of motion defined by the action $S_0[x]$ to give a solution $\gamma$, then we have the next order equations of motion that result from $S_0[x]+\frac{1}{\Lambda^2}S_1[x]$. To solve this next order we solve,
\begin{equation}\label{3.11}
    \frac{\delta S_0}{\delta x}=\frac{1}{\Lambda^2}\frac{\delta S_1}{\delta x}\bigg|_{\gamma}.
\end{equation}
We always evaluate the higher order terms on the previous order's mass as this ensures we only retain a fixed number of modes in the spectrum at any order. Explicitly, following from \eqref{3.11}, if the corrected mass from $S_1$ is given by $\gamma_1$ then our $\Lambda^{-4}S_2$ equation to solve will be $\delta_xS_0=[\Lambda^{-2}\delta_xS_1+\Lambda^{-4}\delta_xS_2]|_{\gamma_1}$. This ensures that no new modes, originating from higher derivative terms, enter the spectrum.

Now for the spin-2 mode, at $\mathcal{O}(1)$ we shall simply obtain a constant mass,
\begin{equation}\label{3.12}
    (-k^2M_P^2+\alpha_2k^4)\Bar{h}_{\mu\nu}=0\rightarrow k^2\bigg(k^2-\frac{M_P^2}{\alpha_2}\bigg)\Bar{h}_{\mu\nu}=0,
\end{equation}
and hence we have a massless spin-2 mode which corresponds to the usual GR result and a spin-2 massive ghost mode with mass,
\begin{equation}\label{3.13}
    M^2=-\frac{M_P^2}{\alpha_2},
\end{equation}
which matches the results found in \cite{StelleHighDeriv}. If $\alpha_2>0$ then this becomes a tachyonic instability, something which is arguably worse than just the normal ghost instability. It is possible to get rid of the ghost by sending $\alpha_2\rightarrow0$ since this sends the ghost mass to infinity and hence it can never be excited. Alternatively, we will see that in this limit the cut-off mass tends to infinity. In order to see this is a ghost for all values of $\alpha_2$ we note the argument outlined in \cite{massivemasslessandghost}; the propagator of the spin-2 mode $\bar{h}_{\mu\nu}$ will be of the form,
\begin{equation}\label{3.14}
    G(k)\propto\frac{1}{k^2}-\frac{1}{k^2+M^2},
\end{equation}
and so the second term has the opposite sign. This clearly marks this mode as a ghost mode in line with the literature \cite{massivemasslessandghost,2005PhLB..608..189N,2005JCAP...03..008C,StelleHighDeriv}. In order to avoid a ghost pathology one may set the cut-off mass to be lower than this mass, that is,
\begin{equation}\label{3.15}
    \Lambda^2<\frac{M_P^2}{\alpha_2}.
\end{equation}
Now we can look at the next order, following the methodology outlined in the above. At this order we shall obtain the equation,
\begin{equation}\label{3.16}
    k^2\bigg(-k^4\tfrac{\alpha_4}{\Lambda^2}+k^2\alpha_2-M_P^2\bigg)\Bar{h}_{\mu\nu}=0,
\end{equation}
which implies,
\begin{equation}\label{3.17}
    M^2\approx-\frac{M_P^2}{\alpha_2}\bigg(1+\frac{\alpha_4}{\alpha_2^2}\frac{M_P^2}{\Lambda^2}\bigg).
\end{equation}
Now that we have the mass, at $\mathcal{O}(\Lambda^{-2})$, \eqref{3.17} and we assume that $\Lambda^2<|M^2|$ we then solve the resulting $\Lambda^2$ quadratic from this inequality to find,
\begin{equation}\label{3.18}
    \Lambda^2<\frac{M_P^2}{2\alpha_2}\Bigg(1+\sqrt{1+4\frac{\alpha_4}{\alpha_2}}\Bigg),
\end{equation}
which raises the cut-off mass if $\alpha_4/\alpha_2>0$. Hence in order for us to make good use of the EFT, that is to raise the cut-off of the theory to explore higher energies, we ought to satisfy $\alpha_4/\alpha_2>0$. Repeating the same procedure for the scalar $\theta$ we find,
\begin{equation}\label{3.19}
    \mathcal{M}^2=\frac{M_P^2}{(6\alpha_1+2\alpha_2)}\bigg(1-\frac{M_P^2}{\Lambda^2}\frac{(6\alpha_3+2\alpha_4)}{(6\alpha_1+2\alpha_2)^2}\bigg).
\end{equation}
Since we are only concerned with eliminating ghost modes we can take \eqref{3.18} as the bound for the cut-off. This bound was raised by the higher order operators we added in the EFT, and it is a general feature that adding further operators would serve to increase this bound. As a result of the above analysis, we see that Quadratic gravity was only a sensible approximation up to around $M_P^2/\alpha_2$. However, with the addition of the cubic terms (and assuming $\alpha_4/\alpha_2>0$) we have been able to raise the cut-off of the theory perturbatively. By adding further terms we would be able to keep raising the cut-off mass by smaller and smaller amounts, with the implicit assumption this infinite sum will tend to infinity in the UV-complete theory.

\subsection{The Case of $v\neq0$}
In the case of spontaneous symmetry breaking of the $\mathbb{Z}_2$ scalar field symmetry, a couple of new features arise which we must deal with before we can calculate the mode masses. Firstly we shall have to deal with terms that appear to be mass terms for the $h$ field, and secondly we shall come across a coupling between $h$ and $\chi$ at quadratic order. The first of these issues requires us to consider a cosmological constant, which we add by hand, and can rid us of these mass-like terms. The second of these problems is much more complex to resolve and since it is not important to the arguments in this paper we shall ignore the spin-0 mode masses.

The cosmological term is straightforward to deal with as outlined above, however the $\mathcal{O}(\Lambda^{-2})$ perturbed action is rather complex and will confuse matters. Hence, we choose to deal with the $\mathcal{O}(1)$ action in order to demonstrate our procedure before we move to the $\Lambda^{-2}$ action. The $\mathcal{O}(1)$ Lagrangian is given by,
\begin{equation}
    \mathcal{L}= \sqrt{-g}\bigg[(M_P^2+\xi_1\phi^2)R-\frac{1}{2}\partial_\mu\phi\partial^\mu\phi+\frac{\mu^2}{2}\phi^2-\frac{\lambda}{4}\phi^4+\alpha_1 R^2+\alpha_2R_{\mu\nu}R^{\mu\nu}-2M_P^2\Gamma_0\bigg]\label{3.20},
\end{equation}
where $\Gamma_0$ is an additional cosmological constant, required to cancel the induced cosmological constant. If we let $\phi=v+\chi$ then the $\phi^2$ and $\phi^4$ terms will contribute a constant in the action and this would typically mean that Minkowski space is no longer a solution to the equations of motion, or if we, a priori, assume that Minkowksi is a solution then we would have a mass term for the spin-2 physical mode \cite{scharf2016gauge}. This is a situation we assume does not occur, that is we assume we always have a massless spin-2 physical mode. However, if we fine tune $\Gamma_0$ then we can get rid of this constant after the SSB has taken place and hence continue to work in Minkoski space. Here, the value of the vacuum expectation value is given by \eqref{2.4} instead of \eqref{2.5}.

The perturbations of the Lagrangian \eqref{3.20} can be found in Appendix \ref{App2.1} which we can now put together. First let us concentrate on the $h$ mass-like term, including the cosmological constant, which we can deal with if we set this term to zero. In doing so we can solve for the cosmological constant which must be given by,
\begin{equation}\label{3.26}
    \Gamma_0=\frac{\mu^4}{8M_P^2\lambda}.
\end{equation}
This means that this theory is \textit{different} to that explored in the previous subsection due to the addition of a cancelling cosmological constant. In order to reproduce the work in the previous subsection we would have to set both $v$ and the $\Gamma_0$ to 0. If we wish to keep a massless spin-2 mode without adding the constant then we have to expand around a curved background. In Section \ref{S4} we shall assume a curved background solution, re-derive the scalar expectation value and then use this to derive the curvature of spacetime. However, for the case at hand the quadratic action is given by,
\begin{equation}\label{3.27}
\begin{aligned}
\mathcal{S}&=\int d^4x\bigg[\frac{M_P^2+\xi_1v^2}{2}(h_{\mu\nu}\Box         h^{\mu\nu}-h\Box h+2h\partial_{\mu}\partial_{\nu}h^{\mu\nu}+2(\partial_\sigma h^{\sigma\mu})^2)\\
    &\qquad+2\xi_1v\chi(\partial_\mu\partial_\nu h^{\mu\nu}-\Box h)+2\alpha_1\bigg(\partial_{\mu}\partial_{\nu}h^{\mu\nu}(\partial_{\mu}\partial_{\nu}h^{\mu\nu}-2\Box h)+h\Box^2h\bigg)\\
    &\qquad+\frac{\alpha_2}{2}\bigg(h_{\mu\nu}\Box^2h^{\mu\nu}+h\Box^2h+2\partial_{\mu}\partial_{\nu}h^{\mu\nu}(\partial_{\mu}\partial_{\nu}h^{\mu\nu}-\Box h)+2\partial_\rho h^{\rho\mu}\Box\partial^\sigma h_{\sigma\mu}\bigg)\\
    &\qquad -\partial_\mu\chi\partial^\mu\chi-(3\lambda v^2-\mu^2)\chi^2\bigg].
\end{aligned}
\end{equation}
Given this we can substitute \eqref{3.7} in \eqref{3.27} and then apply the previous gauge conditions and focus only on the spin-2 ghost mode action. In doing so we obtain a spin-2 action similar to \eqref{3.8} to $\mathcal{O}(1)$,
\begin{equation}\label{3.28}
    \begin{aligned}
    \mathcal{S}_{spin-2}&=\int d^4x\bigg[\frac{(M_P^2+\xi_1v^2)}{2}\Bar{h}_{\mu\nu}\Box\Bar{h}^{\mu\nu}+\frac{\alpha_2}{2}\Bar{h}_{\mu\nu}\Box^2\Bar{h}^{\mu\nu}\bigg],
    \end{aligned}
\end{equation}
from which, noting our previous method, we can simply read off the mass of the spin-2 mode. For the spin-2 ghost mode the mass becomes,
\begin{equation}\label{3.29}
    M^2=-\frac{M_P^2+\xi_1v^2}{\alpha_2}\rightarrow\Lambda^2<\frac{M_P^2}{\alpha_2}+\frac{\xi_1\mu^2}{\alpha_2\lambda},
\end{equation}
hence provided $\xi_1/\lambda>0$ we can increase the cut-off mass via breaking the $\mathbb{Z}_2$ symmetry of the scalar field. We also have the usual spin-2 massless mode that occurs in GR and represents gravitational waves in the theory. It is also interesting to note that the maximum classical cut-off mass can now depend on the scalar bare mode mass, $\mu$, via the scaling relation $\Lambda^2\propto\mu^2$. Furthermore, we can set $M^2\rightarrow\infty$ if we let $\alpha_2\rightarrow\infty$. In doing this we no longer have Quadratic gravity in all generality (rather we have a type of $f(R)$ gravity) and hence renormalizability becomes the issue rather than the ghost.

The main result from this exercise is to show that the cut-off mass can be raised in the case of spontaneous symmetry breaking of the $\mathbb{Z}_2$ scalar field symmetry. We are able to increase the cut-off value as compared to the $v=0$ case, but we pay the price of introducing a cosmological constant hence modifying the action. The spin-2 ghost increase is sourced by the weakly coupled, non-minimal term $\xi_1\phi^2R$ which is assumed to be a small correction to the LEEFT at $\mathcal{O}(1)$ in the effective expansion. During the symmetry breaking, a cosmological constant was induced and in order to retain the massless nature of the spin-2 mode, in flat space, we had to add a preemptive cosmological constant $\Gamma_0$. In the general case, the fine tuning of this cosmological constant relies upon the ratio of $\mu^4/\lambda$. This is an issue that often comes up when breaking matter field symmetries in physics \cite{2007arXiv0708.3184T}. As we shall comment on in Section \ref{S4}, we shall investigate the case in which $\Gamma_0$ is not a fixed constant but is allowed to take on any value. This requires working in the (Anti-)de Sitter background which is a solution to the LEEFT.

Now we shall substitute \eqref{3.1} and \eqref{3.2} into \eqref{2.1} with $v$ given by \eqref{2.5}. Initially we shall just leave the expression as $v$ until we have completed our analysis then we can extract the mass of the fields by transforming into momentum space again. In \eqref{2.1} we had a non-minimal Riemann term of the form $\phi^2R_{\mu\nu\rho\sigma}R^{\mu\nu\rho\sigma}$, however in the expansion this term will be of the form $v^2R_{\mu\nu\rho\sigma}R^{\mu\nu\rho\sigma}$. This means we can eliminate this term via the Gauss-Bonnet theorem in four dimensions at quadratic order by rescaling $\xi_4$ and $\xi_5$, which we shall assume has been done in what follows. The action in this case is somewhat more complex in terms of the coupling constants, however the general structure remains the same as in \eqref{3.27},
\begin{equation}
    \begin{aligned}
    \mathcal{S}=\int d^4x&\bigg(\frac{M_P^2+\xi_1v^2+\frac{\xi_6v^4}{\Lambda^2}}{2}\bigg)(h_{\mu\nu}\Box h^{\mu\nu}-h\Box h+2h\hat{h}+2\hat{h}^\nu\hat{h}_{\nu})\\
    &+\bigg(1+\frac{2v^2(\beta_1+3\beta_2)}{\Lambda^2}\bigg)\chi\Box\chi+\bigg(\mu^2-\frac{30v^4\beta_4}{\Lambda^2}\bigg)\chi^2-\frac{6v^5\beta_4}{\Lambda^2}h\chi\\
    &+\frac{2\beta_3}{\Lambda^2}\chi\Box^2\chi+\bigg(2\alpha_1+\frac{2\xi_4v^2}{\Lambda^2}\bigg)(\hat{h}(\hat{h}-2\Box h)+h\Box^2h)\\
    &+\frac{1}{2}\bigg(\alpha_2+\frac{\xi_5v^2}{\Lambda^2}\bigg)(h_{\mu\nu}\Box^2h^{\mu\nu}+h\Box^2h+2\hat{h}(\hat{h}-\Box h)+2\hat{h}^\nu\hat{h}_\nu)\\
    &+\frac{2\alpha_3}{\Lambda^2}\bigg(\Box\hat{h}(\hat{h}-2\Box h)-h\Box^3h\bigg)+\frac{\alpha_4}{2\Lambda^2}\bigg(2\Box\hat{h}(\hat{h}-\Box h)+2\hat{h}^\nu\Box^2\hat{h}_\nu\\
    &+h\Box^3h+h_{\mu\nu}\Box^3h^{\mu\nu}\bigg)+\bigg(2\xi_1v+\frac{8v^3\xi_6}{\Lambda^2}\bigg)\chi(\hat{h}-\Box h)\\
    &+\bigg(\mu^2v^2-\frac{\lambda v^4}{2}+\frac{2v^6\beta_4}{\Lambda^2}\bigg)\bigg(\frac{1}{8}h^2-\frac{1}{4}h_{\mu\nu}h^{\mu\nu}\bigg).
    \end{aligned}\label{3.30}
\end{equation}
In the limit of all relevant couplings tending to zero we recover the result in \eqref{3.27} which is reassuring. From the quadratic action \eqref{3.30} we can immediately read off the induced cosmological constant required to cancel the last line of \eqref{3.30}. This value is given by,
\begin{equation}
    \Gamma_0=\frac{\mu^4}{8M_P^2\lambda}\bigg(1+\frac{2\mu^2\beta_4}{\Lambda^2\lambda^2}\bigg)\label{3.31},
\end{equation}
which is a marginally larger cosmological constant than in the $\mathcal{O}(1)$ case. We also note that there is a non-zero $h\chi$ coupling, which deviates from the previous example, as a result of the $\xi_i$ terms. One could fine tune these constants to prevent the couplings all together by letting $\xi_1=-4v^2\xi_6/\Lambda^{-2}$, but we do not do that here as we wish to avoid fine tuning specific constants. The other coupled term, proportional to $\beta_4$, would then act as a usual mass term that would need to be diagonalized. By following previous examples we note that the only terms that contribute to the spin-2 mode's mass are the terms of the form $h_{\mu\nu}\hat{\mathcal{O}}h^{\mu\nu}$ where the operator is a scalar operator and hence a function of $\Box$. As such, we can write the $\Bar{h}_{\mu\nu}$ equation down immediately and then solve for the mass. The equation of motion in $k-$space is given by,
\begin{equation}\label{3.32}
    k^2\bigg(-\frac{\alpha_4}{\Lambda^2}k^4+\bigg(\alpha_2+\frac{\xi_5v^2}{\Lambda^2}\bigg)k^2-(M_P^2+\xi_1v^2+\tfrac{\xi_6v^4}{\Lambda^2})\bigg)\Bar{h}_{\mu\nu}=0,
\end{equation}
where we use our previous trick to solve for the mass. That is, we evaluate the $\mathcal{O}(\Lambda^{-2})$ terms on the previous solution and then solve for $k^2$. In doing so, the mass of the spin-2 mode is given by,
\begin{equation}\label{3.33}
    \begin{aligned}
    M^2&=-\frac{1}{\alpha_2+\frac{\xi_5v^2}{\Lambda^2}}\bigg(M_P^2+\xi_1v^2+\frac{\xi_6v^4}{\Lambda^2}+\frac{\alpha_4}{\Lambda^2}\bigg(\frac{M_P^2+\xi_1v^2}{\alpha_2}\bigg)^2\bigg)\\
    &\approx-\frac{1-\frac{\xi_5v^2}{\alpha_2\Lambda^2}}{\alpha_2}\bigg(M_P^2+\xi_1v^2+\frac{\xi_6v^4}{\Lambda^2}+\frac{\alpha_4}{\Lambda^2}\bigg(\frac{M_P^2+\xi_1v^2}{\alpha_2}\bigg)^2\bigg),
    \end{aligned}
\end{equation}
however, $v^2$ is now given by \eqref{2.5} and so we have to substitute this in and expand to order $\Lambda^{-2}$. If we do this, then we obtain,
\begin{equation}\label{3.34}
    \begin{aligned}
    M^2&=-\frac{\lambda M_P^2+\xi_1\mu^2}{\lambda\alpha_2}-\frac{1}{\alpha_2\Lambda^2}\bigg(\frac{\alpha_4}{\alpha_2^2}M_P^4+\frac{2M_P^2\xi_1\alpha_4\mu^2}{\alpha_2^2\lambda}+\frac{\xi_1^2\alpha_4\mu^4}{\alpha_2^2\lambda^2}+\frac{\xi_6\mu^4}{\lambda^2}\\
    &\qquad+\frac{6\xi_1\beta_4\mu^4}{\lambda^3}+\frac{\xi_5M_P^2\mu^2}{\alpha_2}+\frac{\xi_1\xi_5\mu^4}{\alpha_2\lambda^2}\bigg),
    \end{aligned}
\end{equation}
which is quite a lot more complicated that previous expressions. We can then find the classical maximum cut-off by solving $\Lambda^2<|M^2(\Lambda^2)|$, and in doing so we find,
\begin{equation}\label{3.35}
    \Lambda^2\lessapprox\frac{M_P^2}{2\alpha_2}\bigg(1+\sqrt{1+4\frac{\alpha_4}{\alpha_2}}\bigg)+\frac{\mu^2\xi_1}{2\lambda\alpha_2}+\frac{\mu^2\xi_1}{2\alpha_2}\sqrt{1+4\frac{\alpha_4}{\alpha_2}}+\frac{\xi_5\mu^2}{(\alpha_2+4\alpha_4)}\sqrt{1+4\frac{\alpha_4}{\alpha_2}},
\end{equation}
which can raise the cut-off compared to the non-SSB case \eqref{3.17}. Note that \eqref{3.35} is truncated to $\mathcal{O}(1)$, in an $M_P^2$ expansion, as the full solution for $\Lambda$ is rather long and does not shed much light on the raising of the cut-off compared to \eqref{3.35}. When compared to the non-SSB case, we can see that there are a number of sub-leading terms which act to change the cut-off. In order to ensure the cut-off is raised when compared to the previous value, there will be constraints on the $\mathcal{O}(1)$ terms in the expression. Despite these restrictions, we have shown that in the classical case and making some assumptions about the nature of the ghost mode, we can raise the cut-off of the classical effective field theory by including higher order effective terms and breaking the $\mathbb{Z}_2$ symmetry of the scalar field. 

This subsection has shown that the breaking of the $\mathbb{Z}_2$ symmetry of the most general non-minimally coupled EFT expansion of Quadratic gravity allows us to increase the upper-most bound of the classical cut-off. The increase is of subleading order in factors of $M_P^2$ and so does not represent a large increase of the cut-off, however it is sufficient to demonstrate that breaking the symmetry can increase the cut-off. Unfortunately, breaking the symmetry induces a cosmological constant which is required in order to retain the massless nature of the spin-2 physical mode. Hence in the case of keeping $\mathbb{Z}_2$ symmetry, the theory must have a cosmological constant to keep this property.

\subsection{Other EFTs}
So far we have only discussed the full non-minimal Quadratic gravity EFT. We have said nothing about taking Quadratic gravity alone as the LEEFT and adding a non-minimal EFT, and we also have not said anything about the EFT \textit{not} including non-minimal terms. We shall make some comments on these now.

The EFT we have used throughout this work can be split into three sub-theories. The first is the full non-minimal EFT for which we have found the previous results. However, we can consider the case of the EFT without non-minimal terms by setting $\xi_i=0$ in our previous results. In this case the mass of the spin-2 mode is unchanged from the $\mathbb{Z}_2$ symmetry case in \eqref{3.17} to \eqref{3.29} in the SSB case. In fact, at quadratic order the scalar $\chi$ and the gravitational sector decouple completely and so the symmetry breaking will only influence the scalar mode. In this case it is possible to define the masses of the spin-0 fields as there is no longer a direct coupling between the fields which cannot be diagonalized. In this case, the mass of the spin-0 gravitational mode is given by the first order part of \eqref{3.18}. This is not surprising since there are no spin-0 terms in \eqref{3.27} proportional to $v$. The scalar field's mass does change and it becomes $m_\chi^2=2\mu^2$, in line with the typical mass change of scalar field $\mathbb{Z}_2$ symmetry breaking \cite{PhysRevLett.13.508}.

In the case of allowing non-minimal couplings at the EFT expansion level, and not in the LEEFT, the situation for the spin-0 modes in the theory is slightly different as there are genuine couplings. Furthermore, the nature of some of the EFT non-minimal terms means that the vacuum expectation value of the scalar multiplies into $h$ terms and hence contributes to the spin-2 mass. This in turn means that the difference between the two upper-most cut-offs will differ by terms proportional to $\mu^2$.

The point here is that by taking the most general non-minimal EFT expansion we are able to raise the cut-off of the gravitational theory we are dealing with. We have also shown that by breaking the symmetry of the non-minimally coupled theory we are able to raise the cut-off mass even further. These mass differences are not exact, since in the SSB case we had to add an extra cosmological constant on to the theory. However, they are instructive and do demonstrate our ability to change the cut-off mass by going to higher order terms in the effective expansion of the theory. 

\section{Constantly Curved Space}\label{S4}
In the previous section we fixed $\Gamma_0$ so that we could continue to work in flat space without giving the spin-2 mode a mass \cite{scharf2016gauge}, however in this section we shall allow $\Gamma_0$ to be a general constant. This means that flat space is no longer a solution to the equations of motion generated by \eqref{3.19} and hence we shall have to find a new background solution about which to expand the action. As a result, we shall be dealing with classical fields on a curved background and we shall have to deal with the subtleties that come along with such a calculation. Before we proceed, however, it is important to note that in the following we shall assume that the vacuum expectation value is a constant and that one can take a small expansion about the constant $v$. This is the same assumption as above, however it is one that need not always be taken \cite{curvedspace3} and relaxing this assumption can lead to interesting physical consequences. However, in doing so one would introduce a second scale to the theory; the scale associated with the cut-off but also a scale associated with the curvature of spacetime resulting from the expansion of the scalar field solution.

To begin with we have to find the spacetime that satisfies the equations of motion from the action \eqref{3.19} after setting $\phi\rightarrow v$. In order to do this we need to re-derive the value for $v$ and find an equation for the curvature of space $R_0$. This will give us two simultaneous equations to solve for $v$ and $R_0$. The gravitational equations are given by,
\begin{equation}\label{4.1}
    \begin{aligned}
    (M_P^2+v^2\xi_1)G^{\mu\nu}&=\frac{\alpha_1}{2}(2\nabla^{(\mu}\nabla^{\nu)}R-4R^{\mu\nu}R+g^{\mu\nu}R^2-g^{\mu\nu}\overline{\Box}R)\\
    &\quad+\frac{\alpha_2}{2}(g^{\mu\nu}R_{\rho\sigma}R^{\rho\sigma}+2\alpha_2\nabla_\rho\nabla^{(\mu}R^{\nu)\rho}-4\alpha_2g_{\rho\sigma}R^{\mu\rho}R^{\nu\sigma}\\
    &\quad-2\alpha_2\overline{\Box}R^{\mu\nu}-2\alpha_2 g^{\mu\nu}\nabla_\rho\nabla_\sigma R^{\rho\sigma})\\
    &\quad-\frac{1}{8}g^{\mu\nu}(v^4\lambda-2\mu^2v^2+8M_P^2\Gamma_0),
    \end{aligned}
\end{equation}
where $G_{\mu\nu}=R_{\mu\nu}-\frac{1}{2}g_{\mu\nu}R$, $\overline{\Box}=\nabla_\mu\nabla^\mu$, and $(\mu,\nu)=\mu\nu+\nu\mu$. The physical cosmological constant is then given by $\Gamma_{phys}=\frac{1}{8}(v^4\lambda-2\mu^2v^2-8M_P^2\Gamma_0)$. (Anti-)de Sitter spacetime with curvature $R_0$ is actually a solution to this equation of motion provided,
\begin{equation}\label{4.2}
    3R_0(M_P^2+v^2\xi_1)-\frac{1}{8}(8M_P^2\Gamma_0-2\mu^2v^2+v^4\lambda)=0,
\end{equation}
which gives a relation between the vacuum expectation value and the curvature radius. We can also find the scalar equation of motion on this background which gives,
\begin{equation}\label{4.3}
    \lambda v^2=\mu^2+24\xi_1 R_0,
\end{equation}
which we can use in \eqref{4.2} to eliminate $R_0$ and solve for $v^2$. In doing so we have,
\begin{equation}\label{4.4}
    v^2=\frac{\mu^2M_P^2+8\Gamma_0M_P^2\xi_1}{\mu^2\xi_1+\lambda M_P^2}\approx\frac{\mu^2}{\lambda}+\frac{8\Gamma_0\xi_1}{\lambda},
\end{equation}
which recovers the previous result in the limit of $\Gamma_0$ tending to \eqref{3.26}. Now we can put \eqref{4.4} in \eqref{4.3} and solve for $R_0$,
\begin{equation}
    R_0=\frac{8\Gamma_0M_P^2\lambda-\mu^4}{24(\mu^2\xi_1+\lambda M_P^2)}\approx\frac{\Gamma_0}{3}\label{4.5},
\end{equation}
and hence whether we are in de Sitter or Anti-de Sitter space depends largely upon the sign of the cosmological constant. Again, the flat space limit of $R_0=0$ is recovered upon letting $\Gamma_0$ take the value given in \eqref{3.26}. From here we can now perturb the action and then go through and perform the same analysis as we have previously. However, in doing so we shall obtain Klein-Gordon equations with terms that appear to be masses but are in fact simply a result of the fact we are working within a non-flat geometry. For example, in standard General Relativity if we perturb about the constantly curved spacetime then we obtain an equation of the form \cite{scharf2016gauge}, 
\begin{equation}\label{4.6}
    \Box h_{\mu\nu}-2\Gamma_0h_{\mu\nu}=0,
\end{equation}
where the $\Gamma_0$ term is from the cosmological constant. These kinds of terms are \textit{not} mass terms, rather they are a consequence of the cosmological constant/curvature. Hence we shall have to ignore these terms when determining the masses. When we expand about the (Anti-)de Sitter background we shall have non-zero functions for the Riemann and Ricci terms which occur in the quadratic weak-field action. The values for these curvature parameters are well known and they are given by \cite{dSGroup},
\begin{align}
    \Bar{R}_{\mu\nu\rho\sigma}&=R_0(\Bar{g}_{\mu\rho}\Bar{g}_{\nu\sigma}-\Bar{g}_{\mu\sigma}\Bar{g}_{\nu\rho})\label{4.7},\\
    \Bar{R}_{\mu\nu}&=3R_0\Bar{g}_{\mu\nu}\label{4.8},\\
    \Bar{R}&=12R_0\label{4.9},
\end{align}
where from now on we shall use an overbar, $\Bar{O}$, to denote an operator evaluated on the background solution. The calculations, and perturbed action, can be found in Appendix \ref{App2.2}, we do not display them here for brevity. However, it is fairly easy to see that in the limit of $R_0\rightarrow0$, $\cd^\mu\rightarrow\partial^\mu$ and $v^2\rightarrow \mu^2/\lambda$ of \eqref{B14} that we recover the flat space action given in \eqref{3.27}. Now that we have \eqref{B14} in a form that is somewhat familiar, if more complicated, we can begin to apply some gauge condition to get reduce the action to spin-2 and spin-0 actions we have been dealing with throughout this work.

We can apply the previous decomposition into $\bar{h}_{\mu\nu}$ and $\theta$ with the associated gauge conditions provided we note that $\partial_\mu\rightarrow\cd_\mu$ in the curved space. Now as we have seen previously the only terms in the action that contribute to the spin-2 equation of motion are terms of the form $h_{\mu\nu} \Bar{O}h^{\mu\nu}$ where $\Bar{O}$ is some scalar operator. Hence we can read off the spin-2 equation of motion from \eqref{B14}, 
\begin{equation}\label{4.19}
    \bigg(\alpha_2\bar{\Box}^2+(M_P^2+\xi_1v^2+24R_0\alpha_1+2\alpha_2R_0)\bar{\Box}-\frac{1}{4}\Xi\bigg)\bar{h}_{\mu\nu}=0,
\end{equation}
where $\Xi=8M_P^2\Gamma_0+2\mu^2v^2-v^4\lambda-16R_0(M_P^2+\xi_1v^2+6\alpha_1R_0+\alpha_2R_0)$ which are the terms we briefly mentioned in relation to \eqref{4.6}. The first term in $\Xi$ is the contribution from the cosmological constant in the action, and the terms involving $R_0$ and $R_0^2$ are the contributions from the non-flat geometry itself, and hence when we calculate the masses we can ignore this term. In doing so, transforming to $k$ space and then solving for the spin-2 ghost mass we find,
\begin{equation}
    M^2=-\frac{M_P^2+\xi_1v^2+24R_0\alpha_1+2\alpha_2R_0}{\alpha_2}\label{4.20}.
\end{equation}
Now taking the values for $v^2$ and $R_0$, given in \eqref{4.4} and \eqref{4.5} respectively, we find that the limit on $\Lambda^2$ is then given by,
\begin{equation}
    \Lambda^2<\frac{M_P^2}{\alpha_2}+\frac{\mu^2\xi_1}{\alpha_2\lambda}-\frac{(8M_P^2\Gamma_0\lambda-\mu^4)(12\alpha_1\lambda+\alpha_2\lambda+12\xi_1^2)}{12\alpha_2\lambda(M_P^2\lambda+\mu^2\xi_1)},\label{4.21}
\end{equation}
which shows that we can increase the cut-off mass bound if we allow the fraction to be less than 0. Of course there are many ways to allow this to be the case, but the point remains that we are able to raise the cut-off again.

Overall, in this section we have demonstrated that, by allowing the cosmological constant $\Gamma_0$ to take a general value, we are able to raise the cut-off mass bound provided the last term in \eqref{4.21} is less than zero. We can do this in many ways, which involves picking values for $\Gamma_0$, $\mu^2$, $\lambda$, $\alpha_1$ and $\alpha_2$. One way to satisfy this bound is to take $\{\alpha_1,\alpha_2,\lambda,\xi_1\}>0$ and hence we find that $\mu^4>8M_P^2\Gamma_0\lambda$. This leads to the conclusion that,
\begin{equation}
    \Gamma_0<\frac{\mu^4}{8M_P^2\lambda}\label{4.22},
\end{equation}
and hence the cosmological constant must be smaller than its value in the flat space case explored in Section \ref{S3}. In this case, assuming $0<\Gamma_0<\frac{\mu^4}{8M_P^2\lambda}$, and according to \eqref{4.5}, $R_0>0$ and hence we shall be in de Sitter space for the conditions outlined above. It is interesting to note that for $\Gamma_0<0$ and $\Gamma_0>\frac{\mu^4}{8M_P^2\lambda}$ the space is Anti-de Sitter, and hence occupies a much larger proportion of parameter space. In fact from Section \ref{S3}, only one point produces the flat space limit, $\Gamma=\frac{\mu^4}{8M_P^2\lambda}$.

Of course, this is a very specific example and one can choose a variety of points in the parameter space which allows an increase in the upper bound of the cut-off mass. These choices will then determine the nature of the spacetime that one works in, alternatively one could choose the spacetime one wishes to explore, and in turn this will impose conditions on the effective couplings constants.

\section{Conclusion}\label{S5}
In this paper we have used the ghost mode that appears in the mode spectrum of Quadratic gravity to define an approximate cut-off value for the theory if it is treated as a bottom-up effective field theory. We assumed that Quadratic gravity is the low energy effective classical field theory of some unknown UV-completion, and then built our bottom-up theory from there. In order to do this we have made some assumptions about the behaviour of bottom-up effective field theories and ghost modes within their spectrum. The central assumption we have made is, if an EFT contains a ghost mode it is the result of truncating the effective expansion and \textit{not} the result of some pathology within the theory itself. The absolute value of the ghost mode mass squared, that is $|m_g^2|$, is then assumed to be the approximate value of momentum where the effective expansion breaks down. Adding further orders onto the expansion will then increase the absolute value of the ghost mass squared and hence the cut-off. In the infinite limit of the sum of effective operators, which actually defines the unknown UV-completion of the theory, one assumes that the cut-off becomes infinite hence producing a theory that is valid in the UV. In this work we have only worked in the classical limit of a field theory, this is because once one moves into the quantum theory the idea of a cut-off mass becomes more subtle and requires calculation of scattering amplitudes, including loop calculations as well as considerations of perturbative unitarity.

Section \ref{S2} was concerned with setting out the effective action we would be dealing with in Sections \ref{S3} and \ref{S4} and involved writing down all terms that satisfied the symmetries of the problem and that would also only contribute at quadratic order. There were eight other operators that we could have written down at this order which are cubed in the curvature operators \cite{dimsixops}. However in the flat space limit of Section \ref{S3} these operators do not contribute to the quadratic order of the $h$ equations of motion and so we did not include them. In Section \ref{S4} these terms would have contributed but we decided not to investigate this order as it was sufficient to work to $\mathcal{O}(1)$ in the action to demonstrate that curved space allows us to increase the upper bound of the cut-off mass. Furthermore, working with these operators in curved space produces rather complicated equations which do not illuminate the subject any further. We also demonstrated that flat space and a constant $\phi$ value are solutions to the scalar and gravitational equation of motion. Then we derived the vacuum expectation values for the $\mathcal{O}(1)$ and $\mathcal{O}(\Lambda^{-2})$ actions, given by \eqref{2.4} and \eqref{2.5} respectively.

In Section \ref{S3} we worked in the flat space limit in which we preserved the $\mathbb{Z}_2$ symmetry of the scalar field and also in the case where we broke the symmetry. In the case of keeping the $\mathbb{Z}_2$ symmetry, we were able to show that we were able to raise the maximum value to,
\begin{equation}
    \Lambda^2<\frac{M_P^2}{2\alpha_2}\Bigg(1+\sqrt{1+4\frac{\alpha_4}{\alpha_2}}\Bigg)\label{5.1},
\end{equation}
which is greater than the $\mathcal{O}(1)$ value provided $\alpha_4/\alpha_2>0$. This is a general feature of any bottom-up effective field theory. By adding the next order of terms, and putting the relevant bounds on the effective coefficients, we are always able to increase the upper bound of the cut-off value. We assume that in the infinite limit of the sum of effective operators the cut-off becomes infinite and defines the UV-complete theory. Further in Section \ref{S3} we allowed the $\mathbb{Z}_2$ symmetry to be broken, and in order for the Minkowski metric to still be a solution, we had to add a cosmological constant term, $\Gamma_0$. The value of this cosmological constant term then had to be taken as \eqref{3.31} and the cut-off mass' maximum bound was given, approximately, by 
\begin{equation}\label{5.2}
    \Lambda^2\lessapprox\frac{M_P^2}{2\alpha_2}\bigg(1+\sqrt{1+4\frac{\alpha_4}{\alpha_2}}\bigg)+\frac{\mu^2\xi_1}{2\lambda\alpha_2}+\frac{\mu^2\xi_1}{2\alpha_2}\sqrt{1+4\frac{\alpha_4}{\alpha_2}}+\frac{\xi_5\mu^2}{(\alpha_2+4\alpha_4)}\sqrt{1+4\frac{\alpha_4}{\alpha_2}}.
\end{equation}
The approximation is taken as an expansion in $M_P^2$, and this is done as the full equation for $\Lambda^2$ is rather complicated. The raising of the cut-off depends on the sum of the last three terms in \eqref{5.2} being greater than zero, and as in \eqref{4.21} this can be done in a number of ways.

In Section \ref{S4} we allowed the value of $\Gamma_0$ to take on an arbitrary value and hence we were required to work in a general spacetime. However, we were able to show that (Anti) de Sitter space was a solution to the equation of motion and so we were able to expand the perturbations about constantly curved space. In doing so we had to solve \eqref{4.2} and \eqref{4.3} for the vacuum expectation of $\phi$ and the curvature $R_0$ given in \eqref{4.4} and \eqref{4.5} respectively. We were also required to deal with the covariant derivatives associated with the curved background, and in order to manipulate this into a form familiar from the flat space case we had to use \eqref{B11} extensively. After manipulating the initial expressions derived we were able to find the quadratic action given in \eqref{B14}. This action, in the limit of $R_0\rightarrow0$ and $\cd\rightarrow\partial$, becomes the same action as in \eqref{3.27}. Then noting that only terms of the form $h_{\mu\nu}\Bar{O}h^{\mu\nu}$ contribute to the spin-2 equation of motion we were able to write down \eqref{4.19}. However, since we worked in curved space we expected a mass-like term in the spin-2 equation of motion, given by $\Xi$ in \eqref{4.19}. This term did not contribute to the physical mass of the mode since it is a result of the spacetime (and cosmological constant). This led to a ghost mass similar to the flat case but with extra terms dependent on the curvature of the spacetime $R_0$ via \eqref{4.5}. The upper-bond of the cut off mass was then given by,
\begin{equation}
    \Lambda^2<\frac{M_P^2}{\alpha_2}+\frac{\mu^2\xi_1}{\alpha_2\lambda}-\frac{(8M_P^2\Gamma_0\lambda-\mu^4)(12\alpha_1\lambda+\alpha_2\lambda+12\xi_1^2)}{12\alpha_2\lambda(M_P^2\lambda+\mu^2\xi_1)},\label{5.3}
\end{equation}
which demonstrates that the cut-off mass can be raised, even at $\mathcal{O}(1)$, in the constantly curved spacetime provided the final term in \eqref{5.3} is less than zero. 

Overall, this work demonstrates, at least at the classical level, that Quadratic gravity is a safe theory provided it is viewed from a bottom-up effective field theory point of view. From this perspective the pathological spin-2 ghost mode is not a result of issues within the theory but rather our ignorance of the full UV-complete theory. By adding the higher order derivative operators onto the field theory we are able to raise the theoretical cut-off mass of the theory until in the infinite limit we have a full UV-complete theory. We stress that the work carried out here is only valid in the classical limit and when viewing the theory as a quantum field theory greater care must be taken to ascertain the value for the cut-off mass. It would be interesting to repeat this calculation in the quantum limit and determine the cut-off mass in the quantum theory and compare it to the values derived in this work in order to determine if the assumptions we have made here, concerning the spin-2 ghost mass as the cut-off, are accurate. One would have to determine the point at which quantum perturbative unitarity broke down \cite{2017JHEP...09..072D} or use power counting methods \cite{massgraveft}. Working in the classical limit it may also be of interest to explore more general spacetimes such as cosmological spacetimes and black hole geometries to determine the nature of the cut-off in these spacetimes. It would also be interesting to consider solving the scalar field equation of motion exactly using perturbation methods as in \cite{curvedspace3}. Then one could explore the mass of the spin-2 ghost, and hence cut-off, of various spacetimes in a more precise way. This would involve working with two scales in the theory, one being the cut-off and one being the expansion parameter of the scalar field perturbation, which would ultimately involve an expansion in curvature parameters such as the Ricci Scalar. One would have to take great care in this case.

It would also be interesting to consider the FRW case, and consider more concretely the cosmological implications of the non-minimal coupling and the effects SSB of the $\mathbb{Z}_2$ symmetry. There is a large body of work already in this area for Modified Gravity theories \cite{Clifton:2011jh}, our approach would be to treat the effective terms as perturbations from the Quadratic Gravity low energy cosmology. This stream of work will be left for another work.

Finally, it is also interesting to note what happens, if we assume symmetry breaking occurs at some time $T_0$. For example, let us assume that we desire a spacetime that is flat for $t>T_0$ with a massless spin-2 mode. This would imply that we need a pre-emptive cosmological constant to cancel the induced constant produced by symmetry breaking, much as we did in Section \ref{S3}. However, if we assume the $\Gamma_0$ cosmological constant is truly constant in time, this would mean that for $t<T_0$ we would have a non-zero physical cosmological constant given by $\Gamma_0$. Hence our geometry would have to be curved in order to produce a massless spin-2 mode. Thus at the time of symmetry breaking $T_0$, the geometry of the universe would change from (A)dS to Minkowski. The alternative would be to have a flat spacetime for all $t$ and a massive spin-2 mode, \`a la \cite{scharf2016gauge}, in the period $t<T_0$, followed by a massless spin-2 mode in the $t>T_0$ regime. This second interpretation treats the perturbation $h_{\mu\nu}$ like a gauge field propagating on a background.

\section*{Acknowledgements}
CLH would like to thank Douglas Smith at Durham University for his guidance through this project and his incredibly useful conversations. In this work we have extensively used the $xTensor$, $xTras$ \cite{NUTMA20141719} and $xPert$ \cite{2009GReGr..41.2415B} packages from $xAct$ in $Mathematica$.

\newpage

\appendix
\section{Some Extra Notes on Masses}\label{App1}
In this Appendix we make additional comments on the masses of the modes derived in the main body of the paper. We do not look at the coupled situation here as it does not effect the arguments made in this paper.

Firstly, let us show that there is only one spin-0 massive gravitational mode that comes from the LEEFT. The gravitational equations of motion that come from non-minimally coupled Quadratic gravity are given by,
\begin{equation}
    \begin{aligned}\label{A1}
    M_P^2G_{\mu\nu}+T_{\mu\nu}=&2\alpha_1(\nabla_\mu\nabla_\nu-g_{\mu\nu}\Box)R+\frac{\alpha_2}{2}g_{\mu\nu}R_{\rho\sigma}R^{\rho\sigma}\\
    &-2\alpha_2\tensor{R}{^\rho_\mu}R_{\rho\nu}-2\alpha_2(g_{\mu\nu}\nabla_\rho\nabla_\sigma R^{\rho\sigma}+\Box R_{\mu\nu})\\
    &+4\alpha_2(\nabla^\rho\nabla_\nu R_{\rho\mu}+\nabla^\rho\nabla_\mu R_{\rho\nu}),
    \end{aligned}
\end{equation}
where $T_{\mu\nu}=\tfrac{1}{2}\partial_\mu\phi\partial_\nu\phi-\frac{1}{4}g_{\mu\nu}g^{\rho\sigma}\partial_\rho\phi\partial_\sigma-\tfrac{1}{2 }g_{\mu\nu}(\tfrac{1}{2}m^2\phi^2+\tfrac{\lambda}{4}\phi^4)-\xi_1\phi^2G_{\mu\nu}$ is the energy momentum tensor of the scalar field. We can then trace over this equation to give,
\begin{equation}\label{A2}
    (6\alpha_1+2\alpha_2)\Box R=M_P^2R+\tensor{T}{^\mu_\mu}, 
\end{equation}
where $\tensor{T}{^\mu_\mu}$ constitute interactions of the fields. In fact, when the expectation value of the scalar field is 0 as in the first part of Section \ref{S3}, then $\tensor{T}{^\mu_\mu}=0$ and so this term causes no issues. Thus in the flat space, $v=0$, case $X=(6\alpha_1+2\alpha_2)R$ and hence find the spin-0 Klein-Gordon equation of motion given by,
\begin{equation}\label{A3}
    \Box X=\frac{M_P^2}{6\alpha_1+2\alpha_2}X,
\end{equation}
where the mass of the spin-0 mode matches that found in Section \ref{S3}. This spectrum (the type and number of modes) will not change under $\mathbb{Z}_2$ symmetry breaking. Hence we only have a single spin-0 massive mode in the spectrum, and the apparent massless mode, resulting from the extra $\Box$ factor, does not exist. An analogy can be drawn with the Einstein-Hilbert action in which there appears to be a spin-0 massless mode. However, in GR, this mode can be gauged away leaving just the spin-2 massless degree of freedom.

\medskip

We now present a short calculation that results in a recursive formula for perturbatively calculating the spin-2 ghost mode mass in the flat space, $v=0$ limit from Section \ref{S3}. The general spin-2 Lagrangian for such a situation is given by,
\begin{equation}\label{A5}
    \mathcal{L}=\frac{M_P^2}{2}\bar{h}_{\mu\nu}\Box\bar{h}^{\mu\nu}+\frac{\rho_0^2}{2}\bar{h}_{\mu\nu}\Box^2\bar{h}^{\mu\nu}+\frac{1}{2}\sum_{m=1}^\infty\frac{\rho_m}{\Lambda^{2m}}\bar{h}_{\mu\nu}\Box^{(m+2)}\bar{h}^{\mu\nu},
\end{equation}
where $\rho_m$ are unknown effective coupling constants in the theory. As we have seen in Section \ref{S3}, the terms that contribute to the spin-2 ghost action are given by the form $R_{\mu\nu}\Box^nR^{\mu\nu}$, and so these are the terms in the effective action which contribute to the sum in \eqref{A5}. The equations of motion are then given by,
\begin{equation}\label{A6}
    \sum_{m=1}^\infty\frac{\rho_m}{\Lambda^{2m}}\Box^{(m+2)}\bar{h}_{\mu\nu}+\rho_0^2\Box^2\bar{h}_{\mu\nu}+M_P^2\Box\bar{h}_{\mu\nu}=0,
\end{equation}
which we then put into momentum space. This then gives a modified dispersion relation where $k$ must solve,
\begin{equation}\label{A7}
    \sum_{m=1}^\infty\frac{\rho_m}{\Lambda^{2m}}\frac{(-1)^m}{\rho_0^2}k^{(2m+2)}+k^2-\frac{M_P^2}{\rho_0^2}=0,
\end{equation}
however this must be solved perturbatively and hence this defines a recursion relation for solving the mass. This recursion relation is then given by,
\begin{equation}\label{A8}
    m_{N+1}^2=-\frac{M_P^2}{\rho_0^2}\bigg(1+\sum_{n=1}^N\frac{m_N^{(2n+2)}}{M_P^2}\frac{\rho_n}{\Lambda^{2n}}(-1)^{(n+1)}\bigg),
\end{equation}
where $m_1=-M_P^2/\rho_0^2$. The third order mass, corresponding to dimension-8 operators, is then given by,
\begin{equation}\label{A9}
    |m^2_{(3)}|=\frac{M_P^2}{\rho_0^2}\Bigg(1+\frac{M_P^2}{\Lambda^2}\frac{\rho_1}{\rho_0^2}+\frac{M_P^4}{\Lambda^4}\bigg(\frac{2\rho_1+\rho_2\rho_0}{\rho_0^4}\bigg)\Bigg),
\end{equation}
which will increase the maximum classical cut-off value again since we are able to increase the `mass' of the spin-2 ghost by choosing the values of $\rho_m$ carefully.

\section{Deriving the Effective Action}\label{App2}
Here we give some of the calculations used in the main body of the paper.

\subsection{Flat Space}\label{App2.1}

Here we shall deal with the expansion of \eqref{3.20} term by term, expanding to only second order in the fields, as above. Dealing with the $R$ term we have,
\begin{equation}\label{B1}
    \begin{aligned}
    \sqrt{-g}(M_P^2+\xi_1\phi^2)R&=\frac{1}{2}(M_P^2+\xi_1v^2+2\xi_1v\chi)(h_{\mu\nu}\Box h^{\mu\nu}-h\Box h+2h\partial_{\mu}\partial_{\nu}h^{\mu\nu}\\
    &\qquad+2(\partial_\sigma h^{\sigma\mu})^2+2(\partial_{\mu}\partial_{\nu}h^{\mu\nu}-\Box h))\\
    &=\frac{\xi_1v^2+M_P^2}{2}(h_{\mu\nu}\Box h^{\mu\nu}-h\Box h+2h\partial_{\mu}\partial_{\nu}h^{\mu\nu}+2(\partial_\sigma h^{\sigma\mu})^2)\\
    &\qquad+2\xi_1v\chi(\partial_\mu\partial_\nu h^{\mu\nu}-\Box h),
    \end{aligned}
\end{equation}
where the term on the last line results in the first of the coupled spin-0 field terms we have come across. It is obvious that in the limit of no non-minimal interactions then we obtain a field theory which is decoupled. The scalar field kinetic terms is relatively straight forward and we have,
\begin{equation}\label{B2}
    -\frac{1}{2}\sqrt{-g}\partial_\mu\phi\partial^\mu\phi=-\partial_\mu\chi\partial^\mu\chi.
\end{equation}
The mass term for the scalar field perturbs to second order to give,
\begin{equation}
    \sqrt{-g}\frac{\mu^2}{2}\phi^2=\mu^2\chi^2+\mu^2vh\chi+\mu^2v^2(\tfrac{1}{8}h^2-\tfrac{1}{4}h_{\mu\nu}h^{\mu\nu}),
\end{equation}\label{B3}
where we can now see the introduction of mass-like terms for the spin-0 and spin-2 physical gravitational modes as well as another coupled term. We shall deal with the first of these below. As for the quartic scalar field term we obtain,
\begin{equation}\label{B4}
    -\sqrt{-g}\frac{\lambda}{4}\phi^4=-3\lambda v^2\chi^2-\lambda v^3\chi h-\frac{\lambda v^4}{2}(\tfrac{1}{8}h^2-\tfrac{1}{4}h_{\mu\nu}h^{\mu\nu}),
\end{equation}
which introduces another coupled term and mass-like term. The quadratic terms, $\alpha_1$ and $\alpha_2$, are the same as in \eqref{3.3} and are given by,
\begin{equation}
    \begin{aligned}
    \sqrt{-g}(\alpha_1 R^2+\alpha_2 R_{\mu\nu}R^{\mu\nu})&=2\alpha_1\bigg(\partial_{\mu}\partial_{\nu}h^{\mu\nu}(\partial_{\mu}\partial_{\nu}h^{\mu\nu}-2\Box h)+h\Box^2h\bigg)\\
    &\qquad+\frac{\alpha_2}{2}\bigg(h_{\mu\nu}\Box^2h^{\mu\nu}+h\Box^2h+2\partial_{\mu}\partial_{\nu}h^{\mu\nu}\\
    &\qquad\times(\partial_{\mu}\partial_{\nu}h^{\mu\nu}-\Box h)+2\partial_\rho h^{\rho\mu}\Box\partial^\sigma h_{\sigma\mu}\bigg).
    \end{aligned}\label{B5}
\end{equation}

\subsection{Curved Space}\label{App2.2}

Here we shall get on with perturbing \eqref{3.20} in the (A)dS background using (\ref{4.7}-\ref{4.9}), which we shall do term-by-term to make the calculation more explicit. Note that we shall leave $v$ and $R_0$ in the following equations in order to keep things concise, and where relevant we shall substitute them in. We shall begin with the Einstein-Hilbert action with the cosmological constant which gives,
\begin{equation}
    \begin{aligned}\label{B6}
    \sqrt{-g}M_P^2(R-2\Gamma_0)&=\frac{M_P^2}{2}(h_{\mu\nu}\Bar{\Box}h^{\mu\nu}-h\Bar{\Box}h+2h(\Bar{\nabla}_\mu\Bar{\nabla}_\nu h^{\mu\nu})-2h^{\mu\nu}\Bar{\nabla}_\rho\Bar{\nabla}_\nu\tensor{h}{_\mu^\rho})\\
    &\quad +M_P^2\Gamma_0(h_{\mu\nu}h^{\mu\nu}-\tfrac{1}{2}h^2),
    \end{aligned}
\end{equation}
where we see the structure is almost the same as in \eqref{B1} with the partial derivatives upgraded to covariant derivatives and the inclusion of the $\Gamma_0$ constant. There is some subtlety in the structure due to the fact the covariant derivatives do not commute and this is something we shall have to deal with when it comes to applying our gauge fixing conditions in the action. Next we shall deal with the terms which are quadratic in the curvature parameters, beginning with the $R^2$ term,
\begin{equation}
    \begin{aligned}
    \sqrt{-g}\alpha_1R^2=&2\alpha_1(h\Bar{\Box}^2h-2(\Bar{\nabla}_\mu\Bar{\nabla}_\nu h^{\mu\nu})\Bar{\Box}h+(\Bar{\nabla}_\mu\Bar{\nabla}_\nu h^{\mu\nu})^2+6R_0h(\Bar{\nabla}_\mu\Bar{\nabla}_\nu h^{\mu\nu})\\
    &\quad+6R_0h_{\mu\nu}\Bar{\Box}h^{\mu\nu}-12R_0h^{\mu\nu}\Bar{\nabla}^\rho\Bar{\nabla}_\nu h_{\mu\rho}+18R_0^2(4h_{\mu\nu}h^{\mu\nu}-h^2)),
    \end{aligned}\label{B7}
\end{equation}
and the $R_{\mu\nu}R^{\mu\nu}$ gives something more complex,
\begin{equation}
    \begin{aligned}
    \sqrt{-g}\alpha_2R_{\mu\nu}R^{\mu\nu}&=\frac{\alpha_2}{2}((\Bar{\nabla}_\mu\Bar{\nabla}_\nu h)^2+h_{\mu\nu}\Bar{\Box}^2h^{\mu\nu}+2\cd^\mu\cd^\nu h\bar{\Box}h_{\mu\nu}-4\cd_\mu\cd^\nu h^{\mu\rho}\bar{\Box}h_{\nu\rho}\\
    &\quad-2\cd^\rho\cd^\nu h\cd^\nu \cd_\rho h_{\mu\nu}+2\cd_\mu\cd^\rho h^{\mu\nu}\cd^\sigma\cd_\rho h_{\sigma\nu} \\
    &\quad-2\cd^\mu\cd^\nu h\cd^\rho\cd_\nu h_{\mu\rho} +2\cd_\mu\cd^\rho h^{\mu\nu}\cd^\sigma\cd_\nu h_{\rho\sigma})\\
    &\quad+3\alpha_2R_0(3h_{\mu\nu}\Bar{\Box}h^{\mu\nu}-h\Bar{\Box}h+4h\Bar{\nabla}_\mu\Bar{\nabla}_\nu h^{\mu\nu}-6h_{\mu\nu}\Bar{\nabla}_\rho\Bar{\nabla}^\nu h^{\mu\rho})\\
    &\quad+9\alpha_2R_0^2(4h_{\mu\nu}h^{\mu\nu}-h^2),
    \label{B8}
    \end{aligned}
\end{equation}
both of which will require commuting the covariant derivatives in order to apply the gauging conditions we shall impose later. The scalar part of the action is relatively straight forward to find,
\begin{equation}
    \begin{aligned}
    \sqrt{-g}\bigg[-\frac{1}{2}\partial_\mu\phi\partial^\mu\phi+\frac{\mu^2}{2}\phi^2-\frac{\lambda}{4}\phi^4\bigg]&=\chi\Box\chi+(\mu^2-3v^2\lambda)\chi^2+v(\mu^2-v^2\lambda)h\chi\\
    &\quad+\frac{v^2}{8}(2\mu^2-v^2\lambda)(h_{\mu\nu}h^{\mu\nu}-\tfrac{1}{2}h^2),
    \end{aligned}\label{B9}
\end{equation}
which is essentially the same set of terms we had previously in Section \ref{S3}. Now finally we can move on to the non-minimally coupled term which gives,
\begin{equation}
    \begin{aligned}
    \sqrt{-g}\phi^2R&=\frac{\xi_1v^2}{2}(h_{\mu\nu}\Bar{\Box}h^{\mu\nu}-h\Bar{\Box}h+2h\Bar{\nabla}_\mu\Bar{\nabla}_\nu h^{\mu\nu}-2h^{\mu\nu}\Bar{\nabla}^\rho\Bar{\nabla}_\nu h_{\mu\rho})\\
    &\quad+4\xi_1v\chi(\Bar{\nabla}_\mu\Bar{\nabla}_\nu h^{\mu\nu}-\Bar{\Box}h)+12\xi_1R_0(3\chi^2+vh\chi)\label{B10},
    \end{aligned}
\end{equation}
where the first line matches what we expect from \eqref{B6}, and the second line consists of a coupled term we have seen before and another coupled term which is new, that is the $12\xi_1R_0vh\chi$ term. Now, in order to preempt what is to come we could just commute some of the covariant derivatives in the above terms so that we always have at least one covariant derivative, which is contracted into $h^{\mu\nu}$ as the first derivative acting on the object. That is we want the term to be of the form $\Bar{\nabla}_\rho h^{\rho\mu}$. This ensures that these terms become zero when we impose the gauge conditions. However, we shall go further than this and manipulate (\ref{B6}-\ref{B10}) into a form resembling \eqref{3.27}. In order to do this we note the commutator of the covariant derivatives acting on a rank-2 symmetric tensor,
\begin{equation}
    [\nabla_\delta,\nabla_\gamma]T_{\beta\alpha}=\tensor{R}{^\sigma_{\beta\gamma\delta}}T_{\sigma\alpha}+\tensor{R}{^\sigma_{\alpha\gamma\delta}}T_{\beta\sigma}\label{B11}
\end{equation}
and so on for higher rank tensors. To begin with we shall consider the basic example of the term in \eqref{B6},
\begin{equation}
    \begin{aligned}
    \Bar{\nabla}_\rho\Bar{\nabla}_\nu\tensor{h}{_\mu^\rho}&=\Bar{\nabla}_\nu\Bar{\nabla}^\rho h_{\mu\rho}+\Bar{R}_{\sigma\mu\nu\rho}h^{\sigma\rho}+\Bar{R}_{\sigma\nu}\tensor{h}{_\mu^\sigma}\\
    &=\Bar{\nabla}_\nu\Bar{\nabla}^\rho h_{\mu\rho}+R_0(4h_{\mu\nu}-\Bar{g}_{\mu\nu}h)\label{B12},
    \end{aligned}
\end{equation}
which is reassuring as it replicates one of the tensor structures in \eqref{B7} and \eqref{B8}. All except one of the terms \eqref{B7} and \eqref{B8} are of the form $\Bar{\nabla}_\nu\Bar{\nabla}_\rho\tensor{h}{_\mu^\rho}$. The only term not of this form is the $\Bar{\nabla}_\mu\Bar{\nabla}_\nu\Bar{\Box}h^{\mu\nu}$ which requires us to commute the $\Bar{\nabla}_\nu$ derivative past the $\Bar{\Box}=\Bar{\nabla}_\rho\Bar{\nabla}^\rho$ term. We can deal with this in much the same way,
\begin{equation}
    \begin{aligned}
    \Bar{\nabla}_\mu\Bar{\nabla}_\nu\Bar{\Box} h^{\mu\nu}&=\cd_\mu\cd_\rho\cd_\nu\cd^\rho h^{\mu\nu}-\bar{R}_{\sigma\nu}\cd_\mu\cd^\sigma h^{\mu\nu}+\bar{R}_{\sigma\mu\rho\nu}\cd_\mu\cd^\rho h^{\mu\sigma}+\bar{R}_{\sigma\rho}\cd_\mu\cd^\rho h^{\sigma\mu}\\
    &=\cd_\mu\cd_\rho\cd_\nu\cd_\rho h^{\mu\nu}+R_0\cd_\mu\cd_\nu h^{\mu\nu}-R_0\bar{\Box}h\\
    &=\cd_\mu\bar{\Box}\cd_\nu h^{\mu\nu}+5R_0\cd_\mu\cd_\nu h^{\mu\nu}-2R_0\bar{\Box}h\label{B13},
    \end{aligned}
\end{equation}
which gives us precisely what we want. We can carry on with this kind of process, repeating the commutation of covariant derivatives and doing so will take a long while but produces something familiar.

Now we can write down the quadratic action that results from expanding \eqref{3.20} around the constantly curved background defined by \eqref{4.5},
\begin{equation}
    \begin{aligned}
    \mathcal{S}&=\int d^4x\bigg[\frac{M_P^2+\xi_1v^2}{2}(h_{\mu\nu}\bar{\Box}h^{\mu\nu}-h\bar{\Box}h+2h\hat{h}+2\hat{h}^\nu\hat{h}_\nu)+\chi\bar{\Box}\chi+(\mu^2-3v^2\lambda)\chi^2\\
    &\quad +v(\mu^2-v^2\lambda)h\chi+2\alpha_1(h\bar{\Box}^2h-2\hhat\bar{\Box}h+\hhat^2)+\frac{\alpha_2}{2}(h_{\mu\nu}\bar{\Box}^2h^{\mu\nu}+h\bar{\Box}^2h\\
    &\quad-2(\cd_\mu\hhat^\rho)^2+2\hhat^2-2\hhat\bar{\Box}h)+4\xi_1v\chi(\hhat-\bar{\Box}h)+12R_0\alpha_1(h\hhat+h_{\mu\nu}\bar{\Box}h^{\mu\nu}+2\hhat^\nu\hhat_\nu)\\
    &\quad+\frac{\alpha_2R_0}{2}(2h_{\mu\nu}\bar{\Box}h^{\mu\nu}+h\bar{\Box}h+12h\hhat+18\hhat^\nu\hhat_\nu)+12\xi_1R_0(3\chi^2+vh\chi)\\
    &\quad +\frac{1}{8}(8M_P^2\Gamma_0+2\mu^2v^2-v^4\lambda)(h_{\mu\nu} h^{\mu\nu}-\tfrac{1}{2}h^2)\\
    &\quad-R_0(M_P^2+\xi_1v^2+6\alpha_1R_0+\alpha_2R_0)(4h_{\mu\nu}h^{\mu\nu}-h^2)\bigg].\label{B14}
    \end{aligned}
\end{equation}

\newpage
\bibliography{bibliography}
\end{document}